\documentclass[aps,floatfix,twocolumn]{revtex4}
\usepackage{graphicx}
\usepackage{color}

\begin{document}

\title{
Optimal Langevin modelling of out-of-equilibrium molecular dynamics simulations}
 \author{Cristian Micheletti$^1$, Giovanni Bussi$^2$ and Alessandro Laio$^1$ 
\\ $\null$
\\ \small$^1$ International School for Advanced Studies (SISSA) and 
\\ CNR-INFM Democritos, Via Beirut 2-4, 34014 Trieste, Italy
\\ \small$^2$ Computational Science, Dept. of Chemistry and Applied Biosciences, ETH Zurich. 
\\ \small c/o USI Campus, Via Buffi 13, CH-6900 Lugano, Switzerland
}
\date{\today}

\begin{abstract}
We introduce a scheme for deriving an optimally-parametrised 
Langevin dynamics of few collective variables from data generated in
molecular dynamics simulations.  The drift and the position-dependent
diffusion profiles governing the Langevin dynamics 
are expressed as explicit averages over the input
trajectories. The proposed strategy is applicable to cases when the input
trajectories are generated by subjecting the system to a external
time-dependent force (as opposed to canonically-equilibrated
trajectories). Secondly, it provides an explicit control on the
statistical uncertainty of the drift and diffusion profiles. These
features lend to the possibility of designing the external force
driving the system so to maximize the accuracy of the drift and
diffusions profile throughout the phase space of interest.
Quantitative criteria are also provided to assess {\em a
  posteriori} the satisfiability of the requisites for applying the
method, namely the Markovian character of the stochastic dynamics of
the collective variables.
\end{abstract}

\maketitle

With modern molecular dynamics approaches it is possible to follow the
dynamical evolution of systems composed by a large number of
particles. The resulting trajectory, obtained through numerical
integration of the equations of motion, corresponds to a discrete
trace in a phase space of very high dimensionality. In order to
analyze this trajectory it is customary to monitor the time evolution
of only a limited number of collective variables, also referred to as
reaction coordinates, or order parameters. The latter are explicit
functions of the microscopic degrees of freedom of the system for
which they ought to provide a viable coarse-grained description. The
system dynamics and equilibrium properties are then characterized in
terms of these variables alone.

This dimensional reduction strategy, which has ubiquitous applications
in physics and chemistry and biophysics (see
e.g. refs. \cite{garcia92,wham,hirudin,xie1}), has its most-general
formulation in the Zwanzig-Mori projection
procedure\cite{zwanzig}. This scheme is of fundamental conceptual
importance given its general formal applicability to systems whose
evolution is governed by a Liouville operator. In such contexts it can
be demonstrated that the time evolution of the collective variables is
describable by a stochastic dynamics with a non-trivial memory
kernel. Owing to the formidable difficulties posed by the {\em a
priori} determination of the memory kernel, how to devise practical
and general algorithms for carrying out the dimensional reduction
remains an active area of research.

Several approaches have been developed over the years to perform
dimensional reduction in specific
contexts\cite{risken,schulten,grubmuller_diff,xie,hummer,grier,kevrekidis_diff,schutte,grub,kneller2001}. The
validity of the overdamped Langevin dynamics is very commonly assumed
\emph{a priori} for describing the dynamical evolution of the reduced
system.  Consequently, the system free energy landscape and the
diffusion coefficient profile can be expressed in terms of the
Kramers-Moyal coefficients calculated \emph{a posteriori} from
extensive dynamical trajectories \cite{risken}. An interesting
illustration of this strategy is provided in
ref. \cite{kevrekidis_diff} where Kopelevich {\em et al.} estimate the
Langevin drift and diffusion coefficients from short trajectories with
different initial conditions. In other commonly-employed approaches
the Langevin equation parameters are derived from a maximum likelihood
principle\cite{schulten,hummer,schutte}. Specifically, the free energy
and diffusion coefficient profiles are chosen in such a way that the
time evolution of the collective variables actually observed in the
molecular dynamics trajectory has the highest realization
probability. The latter is quantified by computing the Onsager-Machlup
action along the trajectory, see the work of Gullingsrud {\em et al.}
\cite{schulten}. As shown by Hummer \cite{hummer}, this scheme can
also be generalized by allowing for a position-dependent diffusion
coefficient. Another powerful related approach is the one of Horenko
{\em et al.}  \cite{schutte} where the evolution of a system is
assimilated to a diffusive process in a series of harmonic free energy
wells. Transitions between the wells are described as discontinuous
``jump'' processes, with a suitable transition probability per unit
time. The parameters of the model (the position and width of the
harmonic wells, the diffusion coefficients in the wells and the
jumping rates) are also derived {\em a posteriori} from a maximum
likelihood approach.

Most available approaches have been formulated and designed to be
applied to take as input equilibrated (canonical) trajectories. Recent
advances in thermodynamic sampling techniques, however, stimulate the
formulation of more general approaches applicable to systems subjected
to external time-dependent biases. Large systems with a corrugated
energy landscape would spontaneously evolve very slowly and the
introduction of suitable external forces provides an effective means
of driving the system through the reduced phase space. This is
commonly exploited in several thermodynamic sampling techniques, such
as steering\cite{steering}, local-elevation\cite{local_elevation}
adaptive force bias\cite{pohorille}, flooding \cite{flooding},
Wang-Landau\cite{wl} and metadynamics\cite{hills,cristian}. To the
best of our knowledge, the method of Gullingsrud \emph{et al.}
\cite{schulten} is the only available maximum-likelihood approach
which is applicable to systems (whose diffusion coefficient is known
{\em a priori}) subjected to externally-applied biases.

Building on the previous studies mentioned above, we here formulate
and apply a novel maximum-likelihood scheme that allows to recover
efficiently the equilibrium and dynamic properties of the reduced
system even when subjected to an externally-applied time-dependent
force. The variational approach addressed in this study complements
the advantages of the strategies in refs. \cite{hummer} and
\cite{schulten} as it allows recovering {\em a posteriori} a
non-constant diffusion coefficient profile while accounting for
externally-applied forces.

The method provides not only the drift and diffusion coefficients for
the system (in one or more collective variables) but also an estimate
of their statistical errors. For all these quantities we derive
expressions that are straightforwardly calculated by averaging
suitable observables along the dynamical trajectories. The possibility
to control the error on the drift and diffusion terms of the reduced
system opens the possibility to design the applied external bias so to
achieve a pre-assigned profile of statistical uncertainties for the
quantities of interest.

In the following we shall first derive the maximum-likelihood
expressions for the drift and diffusion coefficient profiles and their
errors.  The advantages and range of applicability of the method are
finally illustrated and discussed for a specific system, namely the
problem of looping of a self-avoiding polymer chain in a crowded
medium.

{It should be remarked that the validity of the
approach relies crucially on an appropriate choice of the collective
variable whose dynamics, sampled at appropriate time intervals, must
have a Markovian character. This requirement is not necessarily
fulfilled by an arbitrarily chosen variable.  Indeed, even if the
trajectory of the system is generated by a Markovian process
(e.g. molecular dynamics with Langevin thermostats), the dynamics of a
single, projected variable cannot be expected to be Markovian
too\cite{zwanzig}. Though no simple {\it a priori} criteria can be
adopted for a good choice of collective variables we discuss, in
section \ref{sec:dt}, how quantitative schemes can be introduces for
verifying {\it a posteriori} if a given time series is reliably
described by a Morkovian process and if the overal approach can
consistently be applied.}

\section{Optimal Langevin description of a stochastic dynamics process}

\bigskip We consider a system with several microscopic degrees of freedom
and whose salient properties are described through a much smaller number of
collective variables (CVs), $s_{i=1,...N}$, chosen \emph{a priori} and
defined in terms of the microscopic variables. The system is assumed to
evolve in time under the combined action of two kind of forces: (i) the
thermodynamic force, tending to establish the canonical equilibrium
associated to a given temperature $T$, and (ii) a time-dependent external
force acting on the collective variables. In the following we shall indicate
with $\theta _{i}(t) $ the instantaneous external force conjugated to the $i$th
collective variable. A prototype system, which will be discussed later, is
constituted by a polymer chain where the fundamental degrees of freedom are
the centers of its spherical monomers. A single collective variable will be
used, namely the polymer end-to-end distance. The polymer dynamics is
controlled by both the thermal buffeting of the surrounding solvent
molecules and by an externally-controlled stretching force applied to the
chain ends.

The evolution of the system is followed at the level of the collective
variables through an equispaced time series $\mathcal{T}=\left\{
s\left( 0\right) ,s\left( dt\right) ,\dots ,s\left( t\right) ,\dots
\right\} $, where $s\left( t\right)$ denote the array of the
instantaneous CV's values, $s_{i=1,...,N}\left( t\right) $. The
objective is to take the discrete trajectory $\mathcal{T}$ and the
accompanying time series of the externally-applied forces, $\left\{
\theta \left( 0\right) ,\theta \left( dt\right) ,\dots ,\theta \left(
t\right) ,\dots \right\}$, as the sole inputs for deriving the best
parametrization of the system properties within a Langevin description
of the CV's evolution. In particular, the aim is to recover the
thermodynamic forces and diffusion matrices of the \emph{isolated}
system for a wide range of CV's values by recording how the
\emph{externally-driven} system evolves.

The extraction of the optimal Langevin parametrization is carried out within
a maximum likelihood approach, a framework profitably used in other previous
approaches\cite{schulten,schutte,hummer}. We start by assuming that for a
suitable choice of the discretization time interval, $dt$, the CV's
evolution is describable as a Markovian process. The probability to observe
a specific trajectory $\mathcal{T}$ is accordingly 
\begin{equation}
P[\mathcal{T}]\propto {\prod_{t}}\pi \left( s(t), ds\left( t\right) \right)
\label{eqn:p}
\end{equation}
where $ds_{i}\left( t\right) \equiv s_{i}\left( t+dt\right) -s_{i}\left(
t\right)$ and $\pi \left( s(t), ds\left( t\right) \right)$ is the probability
of the elementary step.

\noindent {For non-externally-driven systems, described by a
single collective variable, $s$, subject to an overdamped Langevin
evolution with constant diffusion coefficient $D$ in a free energy
landscape, $\mathcal{F} (s)$, the probability of the elementary step
has a simple Gaussian form: $\pi \left( s(t), ds\right) \propto
\frac{1}{\sqrt{D}}\exp {\left[ -\frac{1}{4Ddt} \left( ds+\left(
D\partial \mathcal{F}\left( s\right) \right) dt\right) ^{2} \right]}$
\cite{risken}.} For simplicity of notation in the previous expression
and in the following it is implied that the free energy $\mathcal{\
F}$ is expressed in units of the thermal energy, $\kappa _{B}\,T$. For
the case of several collective variables and if the diffusion
coefficients depends on the CV's themselves, the previous expression
generalizes to\cite{risken}:
\begin{equation}
\pi \left( s,ds\right) \propto \frac{1}{\det D\left( s\left( t\right) \right)
^{\frac{1}{2}}}\exp {\left[ -\frac{1}{4dt}D_{ij}^{-1}(s\left( t\right) )\chi
_{i}\left( t\right) \chi _{j}\left( t\right) \right] }  \label{eqn:pi}
\end{equation}
\noindent where a summation of repeated indexes is implied and 
\begin{eqnarray}
\chi _{i}\left( t\right) &=&ds_{i}\left( t\right) +\left( D_{ij}\left(
s\right) \partial _{j}\mathcal{F}\left( s\right)
-\partial_{j}D_{ij}(s)\right) dt \nonumber \\
&=& ds_{i}\left( t\right) -v_{i}\left( s\right) \, dt.  \label{eqn:ds}
\end{eqnarray}
\noindent where $v_i(s)$ is the drift field\cite{risken}.  Without
 loss of generality, the diffusion matrix is assumed to be symmetric:
 $D_{ij}(s)=D_{ji}(s)$\cite{risken}. Eq. (\ref{eqn:p}), (\ref{eqn:pi})
 and (\ref{eqn:ds}), for a given choice of $D\left( s\right) $ and
 $v\left( s\right) ,$ allow computing the probability of a trajectory
 $\mathcal{T}$. \noindent The stochastic differential equation leading
 to Eq. (\ref{eqn:pi}) and (\ref{eqn:ds}) is given by

\begin{equation}
ds_{i}\left( t\right) =v_{i}\left( s\left( t\right) \right) dt+\sqrt{2}\,\,D_{ij}^{1/2}(s\left( t\right) )\,dW_{j}\left( t\right)   \label{langevin}
\end{equation}

\noindent where $\{dW_{i}\left( t\right)\}$ is a $N-$dimensional Wiener
process.

\noindent In the presence of the external force $\theta _{l}\left( t\right)$, Eq. (\ref{eqn:ds}) is modified as follows: 
\begin{eqnarray}
\chi _{i}\left( t\right) &=&ds_{i}\left( t\right) +\left( D_{ij}\left(
s\right) \partial _{j}\mathcal{F}\left( s\right) \right.  \nonumber \\
&&\ \ \ \ \left. -D_{il}\left( s\left( t\right) \right) \theta _{l}\left(
t\right) -\partial _{j}D_{ij}(s)\right) dt  \nonumber \\
&=&ds_{i}\left( t\right) -v_{i}\left( t\right) \,dt-D_{il}\left( s\left(
t\right) \right) \theta _{l}\left( t\right) dt\ .
\end{eqnarray}
The extra term would contribute a term $-D_{ij}\left( s\right) \theta
_{j}\left( t\right) dt$ in Eq. (\ref{langevin}).

\noindent {Eq. (\ref{eqn:p}), supplemented by the
  relations of Eq.  (\ref{eqn:pi}) and (\ref{eqn:ds}), coincides with
  the expression of the Onsager-Machlup action\cite{risken}, namely
  with the probability to observe the trajectory $\mathcal{T}$ given
  the {\em known} diffusion and drift terms governing its langevin
  evolution. In the present context , the probability $P[\mathcal{T}]$
  of Eq. \ref{eqn:p} can also be profitably interpreted from a
  complementary perspective. In fact, for a given trajectory,
  $P[\mathcal{T}]$ can be considered as a likelihood functional which
  characterizes the non-externally-driven system in terms of $v$ and
  $D$ .} In this approach, extremizing Eq. (\ref{eqn:p}) provides the
  {\em unknown} drift and diffusion terms ($v$ and $D$) yielding the
  highest possible probability for the given observed trajectory,
  $\mathcal{T}$.

Notice that, at variance with other approaches\cite{schulten,hummer},
the likelihood of the trajectory is here maximized with respect to the
drift field $v_i(s)$ and not the free energy $F(s)$.  An advantage of
this alternative approach is that the solution can be expressed as an
explicit average computed over the trajectory, at least in the one
dimensional case (see Eq. (\ref{diff_force})).  The disadvantage is
that, when two or more collective variables are used, the optimal $v$
and $D$ are not guaranteed to yield an equilibrium probability
measure\cite{risken}, but only to a stationary one. This condition
should indeed be verified {\it a posteriori} and provides a further
consistency criterion for the viability of the approach.

The cardinal variational equations for the drift and diffusion terms
are $\frac{\delta \log P[\mathcal{T}]}{\delta v_{i}(s)}=0$ and
$\frac{\delta \log P[\mathcal{T}]}{\delta D_{ij}(s)}=0$. After some
algebra one obtains:

\begin{eqnarray}
&&\frac{\delta \log P[\mathcal{T}]}{\delta v_{i}(s)}={\frac{D_{ij}^{-1}(s)}{
2}}\sum_{t}\chi _{j}\left( t\right) \delta _{s-s\left( t\right) }  \nonumber\\
&=& {\frac{ D_{ij}^{-1}}{2}}(s)\sum_{t}\delta_{s-s\left( t\right)}
[ds_{j}-v_{j}dt-D_{ij}\theta _{l}dt]=0 \\
&&\frac{\delta \log P[\mathcal{T}]}{\delta D_{ij}(s)} =\sum_{t}\left[ -\frac{
1}{2}D_{ij}^{-1}(s)+\right.  \nonumber\\
&& \frac{1}{4dt}D_{il}^{-1}\left( s\right) D_{jm}^{-1}(s)\chi _{l}\left(
t\right) \chi _{m}\left( t\right) +\frac{1}{4} D_{il}^{-1}\left( s\right)
\chi _{l}\left( t\right) \theta _{j}\left( t\right)  \nonumber\\
&&\left. +\frac{1}{4}D_{jl}^{-1}\left( s\right) \chi _{l}\left( t\right)
\theta _{i}\left( t\right) \right] \delta _{s-s\left( t\right) }=0
\label{eqn:extr2}
\end{eqnarray}

\noindent where the condition $D_{ij}=D_{ji}$ has been enforced while taking
the variation with respect to $D$. Introducing the notation $\left\langle
a\right\rangle _{s}=\frac{\sum_{t}\delta _{s-s\left( t\right) }a\left(
t\right) }{\sum_{t}\delta _{s-s\left( t\right) }}$, we obtain the following
equations, that, in general, have to be solved self-consistently:

\begin{eqnarray}
v_{i}\left( s\right) &=&\frac{1}{dt}\,\left\langle ds_{i}\right\rangle
_{s}-D_{ij}\left( s\right) \left\langle \theta _{j}\right\rangle _{s}\ .
\label{eqn:solution1} \\
D_{ij}\left( s\right) &=&{\frac{\left\langle ds_{i}ds_{j}\right\rangle
-\left\langle ds_{i}\rangle \langle ds_{j}\right\rangle }{2\,dt}}  \nonumber
\\
&& +{\frac{ D_{ip}(s)D_{kj}(s)}{2}}dt(\left\langle \theta _{p}\theta
_{k}\right\rangle -\left\langle \theta _{p}\rangle \langle \theta
_{k}\right\rangle )  \label{eqn:solution2}
\end{eqnarray}

\noindent Eqs. (\ref{eqn:solution1}) and (\ref{eqn:solution2}) make
 possible to estimate $D\left( s\right) $ and $v\left( s\right) $ also
 from trajectories obtained in the presence of time-dependent forces
 acting on the system. It is noteworthy that
 Eqs.~(\ref{eqn:solution1}) and (\ref{eqn:solution2}) tie the optimal
 estimates of $v$ and $D$ to suitable averages made on the
 trajectory. On one hand this lends to a straightforward numerical
 implementation of the scheme. On the other, it highlights a key
 difference between the Eqs.~(\ref{eqn:solution1}) and
 (\ref{eqn:solution2}) for driven systems and the Kramers-Moyal
 coefficients of first and second order which connect the Langevin and
 Fokker-Planck descriptions of the system evolution. At variance with
 the spirit of the averages in the above equations, in fact, the
 time-dependent Kramers-Moyal coefficients are defined as averages
 over the Wiener process at a given time and value of the CV's.
 {It is, however, clear that for a
 non-externally-driven system, where $\theta (t)=0$ at all times, the
 averages in Eqs.~(\ref{eqn:solution1}) and (\ref{eqn:solution2})
 coincide with those over independent realizations of the noise, and
 hence $v$ and $D$ match the time-\emph{in}dependent Kramers-Moyal
 coefficients:}

\begin{eqnarray*}
v_{i}\left( s\right) &=&\frac{1}{dt}\,\left\langle ds_{i}\right\rangle _{s}
\\
D_{ij}\left( s\right) &=&{\frac{\left\langle ds_{i}ds_{j}\right\rangle
-\left\langle ds_{i}\rangle \langle ds_{j}\right\rangle }{2\,dt}}\ .
\end{eqnarray*}

\noindent In one dimension (namely for $N=1$) Eq. (\ref{eqn:solution2}) can
be explicitly solved also for $\theta \neq 0$. Since $D$ must be 
positive-defined, the second order Eq. (\ref{eqn:solution2}) admits a 
single physically viable solution:

\begin{equation}
D\left( s\right) =\frac{-1+\sqrt{1+\left( \left\langle \theta
^{2}\right\rangle _{s}-\left\langle \theta \right\rangle _{s}^{2}\right)
\left( \left\langle ds^{2}\right\rangle _{s}-\left\langle ds\right\rangle
_{s}^{2}\right) }}{dt\left( \left\langle \theta ^{2}\right\rangle
_{s}-\left\langle \theta \right\rangle _{s}^{2}\right) }
\label{diff_force}
\end{equation}

\noindent 
An important payoff of the maximum-likelihood approach is that it
leads straightforwardly to estimating the uncertainty on the inferred values
of $v_{i}^{\ast }$ and $D_{ij}^{\ast }$ (a star is used to denote the fact
that these values maximize $P[\mathcal{T}]$), associated to the limited
statistics inherent in any trajectory with finite duration. To do so we
consider the expansion of $P[\mathcal{T}]$ around the maximum retaining
terms up to quadratic order. Introducing an $N+N^{2}$ dimensional vector $y\left( s\right) =(\dots ,v_{i}\left( s\right) ,\dots ,D_{ij}\left( s\right)
,\dots )$, the error on $y_{i}$ at $s$, $\sigma ^{2}\left( y_{i}\left(
s\right) \right) $, can be estimated as the standard deviation of ${y_{i}(s)}
$ from its best available estimate, ${y_{i}^{\ast }(s)}$. This is given by $\sigma ^{2}\left( y_{i}\left( s\right) \right) =-\left( A^{-1}\right) _{ii}$
where 
\begin{equation}
A_{ij}\left( s\right) =\left. {\frac{\delta ^{2}\log \left( P[\mathcal{T}]\right) }{\delta y_{i}(s)\delta y_{j}(s)}}\right\vert _{y=y^{\ast }}
\label{eqn:A}
\end{equation}

\noindent\ For $N=1$ we have

\[
A\left( s\right) =-{\frac{N(s)dt}{2D}}\left( 
\begin{array}{cc}
1 & \langle \theta \rangle \\ 
\langle \theta \rangle & {\frac{1}{D\,dt}}[1+dtD\,\langle \theta ^{2}\rangle
]\end{array}
\right) 
\]
\noindent with $N\left( s\right) =\sum_{t}\delta \left( s-s\left( t\right)
\right)$. Thus, the uncertainties on $v$ and $D$ are given by

\begin{eqnarray}
\sigma ^{2}\left( v\left( s\right) \right) &=&\frac{2}{N(s)}\frac{D}{dt} 
\frac{1+dtD\,\langle \theta ^{2}\rangle }{1+dtD(\langle \theta ^{2}\rangle
-\langle \theta \rangle ^{2})} \nonumber \\
\sigma ^{2}\left( D\left( s\right) \right) &=&\frac{2}{N(s)}\frac{D^{2}}{
1+dtD(\langle \theta ^{2}\rangle -\langle \theta \rangle ^{2})}
\label{eqn:errors}
\end{eqnarray}

As intuitively expected, the quadratic error on $v$ and $D$ in $s$ are
inversely proportional to $N\left( s\right)$, that is the number of
times the trajectory visited $s$. For a trajectory generated by an
ordinary dynamics on a system whose reduced free energy exhibits
several minima saddle points etc., this error will be highly
non-uniform. Even if transitions from the various basins are observed,
the statistical accuracy on $v$ and $D$ in the transition region will
degrade very rapidly with the barrier height. This limitations can be
overcome with the aid of expression (\ref{eqn:errors}) and a
preliminary rough knowledge of the free energy profile. In this case
it is conceivable to design the application of the time-dependent
external forces so to achieve an approximately uniform coverage of the
phase space of interest.
 { It should be remarked that Eq.  (\ref{eqn:errors}) quantifies the statistical 
uncertainties on $v$ and $D$ and does not take into account possible systematic errors deriving
from the non-Markovian nature of the process. These aspects will be 
discussed in more detail in Section II.E.}

\section{Application: looping of a polymer chain}

\bigskip In the following we shall discuss the application of the
above strategy to the problem of loop formation in a model polymer
chain fluctuating in a solvent rich in crowding molecules. The polymer
model considered here, follows the one introduced in Ref. \cite{toan}
to study how the polymer looping kinetics is affected by the crowding
agents \cite{maren06}. The novel question addressed here regards the
possibility to describe the evolution of the polymer end-to-end
distance, $s$, by means of a Langevin equation. It will be shown that,
for a suitable choice of the time interval with which the original
trajectory is sampled, a Langevin description for the evolution of $s$
is, in fact, possible. Interestingly, despite the simplicity of the
system and its formulation, both the optimally-recovered drift and
diffusion terms have a non-trivial dependence on $s$.

The model polymer consists of $n$ spherical beads of radius $R$ interacting
via the following potential energy term: 
\begin{equation}
V=\epsilon _{1}\sum_{i<j}e^{-a(d_{i,j}-2R)}-\epsilon _{2}\sum_{i}\ln [1-({\frac{d_{i,i+1}}{1.5\ R}})^{2}]  \label{eqn:Vc}
\end{equation}
\noindent where the $i$ and $j$ denote the sequential indexing of the
$n$ chain beads. The first term in expression (\ref{eqn:Vc}) enforces
the self-avoidance of the chain, while the second provides the
attractive interaction between consecutive beads, thus enforcing the
chain connectivity as in the FENE model\cite{kremer1}. The model
parameters are exactly those introduced in Ref.~\cite{toan} to
describe an eukariotic chromatin fiber, whose effective diameter and
persistence length are both $\sim $ 25 nm
\cite{chromatin}. Specifically, $\epsilon _{1}$ and $\epsilon _{2}$
are respectively 1 and 70 units of thermal energy, $\kappa _{B}T$,
$a=4$ nm$^{-1} $, and $R=12.5$nm is the bead radius. At the chosen
temperature, $T=300$ K, the interplay of the two terms in
(\ref{eqn:Vc}) ensures that distance between consecutive beads
fluctuates around the nominal value of 25 nm by only about 0.5 nm. The
mass of the beads is calculated from the typical densities of
biopolymers, $\rho= 1.35$ g/cm$^3$~\cite{Matthews}.

As anticipated, the motion of the chromatin fiber is assumed to occur
in a medium crowded by other biomolecules (proteins, RNA etc.) which
are simply modeled as monodispersed globular particles of radius
$r=2.5$nm which altogether occupy 15\% of the system volume. The
crowding agents are not modelled explicitly but rather through the
Asakura-Oosawa (AO) mean-field approach \cite{AO}. This approach
exploits the smallness of the crowding agents compared to the chain
beads, to introduce the effective self-attraction of the polymer,
known as depletion interaction, induced by the hard-core repulsion
with the crowding agents. This additional self-interaction is
described by the following potential energy term:
\begin{equation}
V_{AO}=-\frac{\phi k_{B}T}{16r^{3}}\sum_{i<j}\left( 2\tilde{d}_{ij}+3d_{ij}-\frac{3\Delta _{ij}^{2}}{4d_{ij}}\right) \ \tilde{d}_{ij}^{2}\ \Theta
\lbrack \tilde{d}_{ij}]
\end{equation}
\noindent where $\tilde{d}_{ij}=2r+d_{i,j}^{0}-d_{ij}$, $\Delta
_{ij}=\left\vert R_{i}-R_{j}\right\vert $, and the step function
$\Theta $ ensures that the AO depletion interaction vanishes at
distances $>d^{0}+2r$.  The dynamics of each bead (subjected to a
Stokes-Einstein friction appropriate for molecular crowding
\cite{toan}) was followed within a under-damped Langevin
scheme\cite{allen} with an integration time step of 1 ps, appropriate
to resolve the decay of the correlation of the fastest-relaxing
degrees of freedom of the system, the beads velocities.  Part of the
results presented below are obtained analyzing the end-to-end distance
with a sampling time intervals of 200 ps or larger. The evolution of
this quantity occurs over a time scale much larger
than the relaxation time of the beads velocities and hence, for
reasons of efficiency, was obtained through the over-damped Langevin
scheme\cite{allen} with an integration time step of 15 ps. The
equivalence of the {\em under-} and {\em over-}damped schemes was
explicitly verified by comparing the estimates of the drift and
diffusion coefficients obtained by processing runs covering 15 ms.

We first followed the evolution of the isolated system and used the
recorded trajectory for the analysis presented in the previous
section. We considered a single collective variable, namely the
end-to-end distance, $s\left( r\right) =\left\Vert
r_{N}-r_{1}\right\Vert $.  {As anticipated in the
introduction, although the evolution of the original $n$-particle system
is Markovian, the dynamics of $s\left( r\right)$ might not be
necessarily so. This issue will be discussed in detail in Section
\ref{sec:dt}.}  We set the chain length, $n$, equal to 5; this made
possible to collect extensive trajectories with an affordable
computational effort and hence validate, at least in part, the
variational Langevin description by comparing the predicted
equilibrium properties against data obtained by straightforward
histogram techniques.

\subsection{Equilibrium properties of the optimal model}

\begin{figure}
\includegraphics*[width=0.45\textwidth]{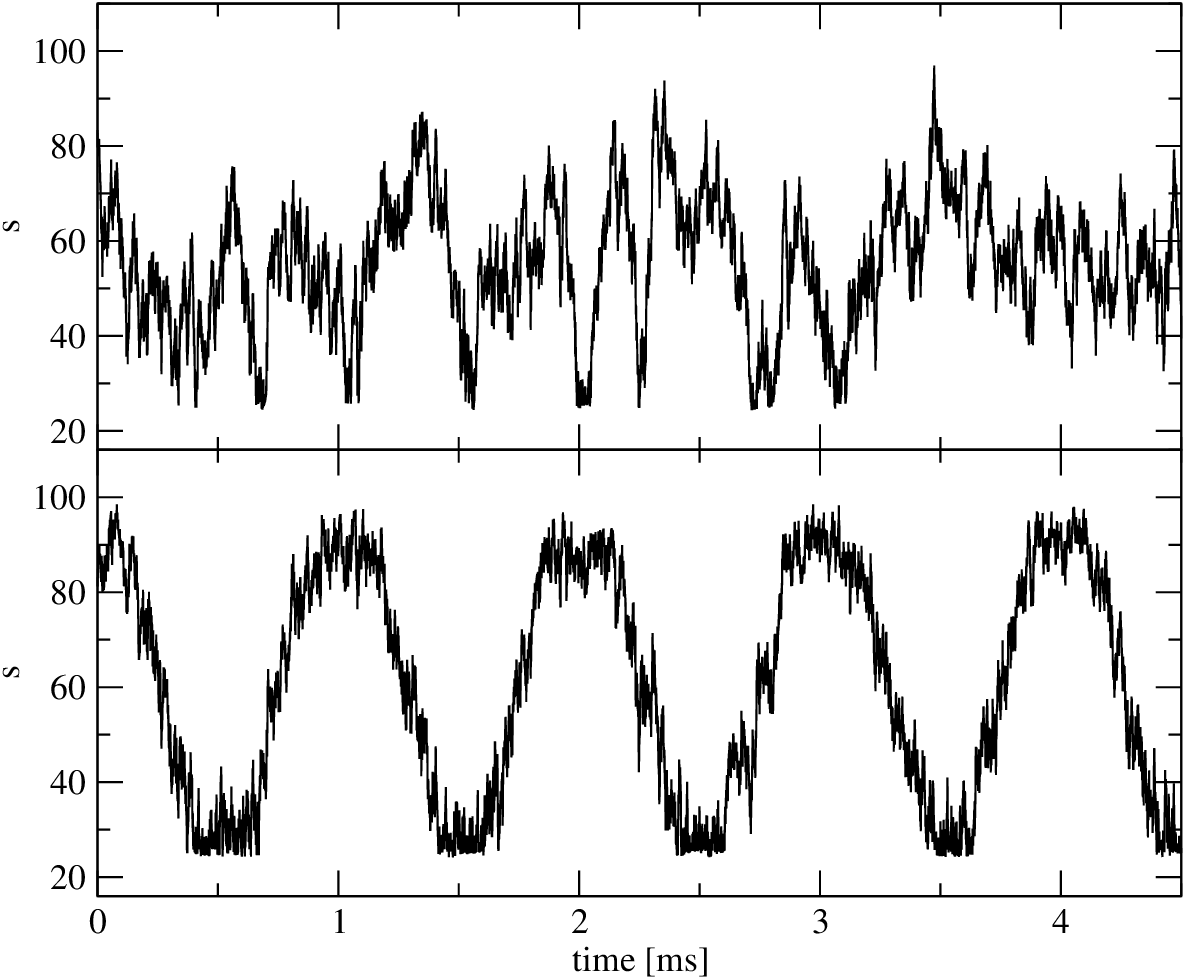}
\caption{Evolution of the end-to-end distance in the absence (upper panel)
and presence (lower panel) of an external force.}
\label{fig:traj}
\end{figure}

Starting from a random configuration of the polymer we have initially
followed its underdamped Langevin dynamics (in the absence of any
external force) over a time span of 180ms.
  As illustrated by the sample time series of $s$, shown in
Fig. \ref{fig:traj}, upper panel, this time span is much larger than
the typical looping/unlooping times of the chain, and hence is a
sufficient guarantee of equilibration of the system properties. This
trajectory was consequently used to estimate $v\left( s\right) $ and
$D\left( s\right) $ and their errors from the averages in Eqs.
(\ref{eqn:solution1}), (\ref{eqn:solution2}) and (\ref{eqn:errors}). In
order to compute these averages it is necessary to choose the value
for $dt$ entering in the Langevin equation. The correct $dt$ has to
satisfy two conditions.  First of all, $dt$ must be so large that
the underlying process can be considered, at least approximately, as
Markovian. At the same time, $dt$ has to be sufficiently small that
the typical change of reaction coordinate over such time scale does not
reflect in a significant variation of the free energy and diffusion
coefficient. 

This second condition can in principle be relaxed if the transition
probability of the underlying Langevin process was known for an
arbitrary $dt$ \cite{schutte}. Exact expressions for finite-time
transition probabilities are however available for very special
potentials, notably harmonic ones \cite{schutte}. In the present case,
the free energy shape is not specified {\em a priori}, and hence it is
necessary to use the approximation of Eq. (\ref{eqn:pi}), which is valid
for small $dt$.  Thus, the two conditions specified above may be
potentially mutually exclusive.

For the simple polymer model described in this
work it is possible to compute accurate equilibrium and kinetic
quantities directly from extensive simulations, and choose {\em a
posteriori} the value of $dt$ in order to obtain a Langevin model 
that reproduces them in as faithfully as possible.
An alternative practical criterion for choosing this parameter without
benchmarking the Langevin predictions against the exact results will
be described in the following.

\begin{figure}
\includegraphics*[width=0.45\textwidth]{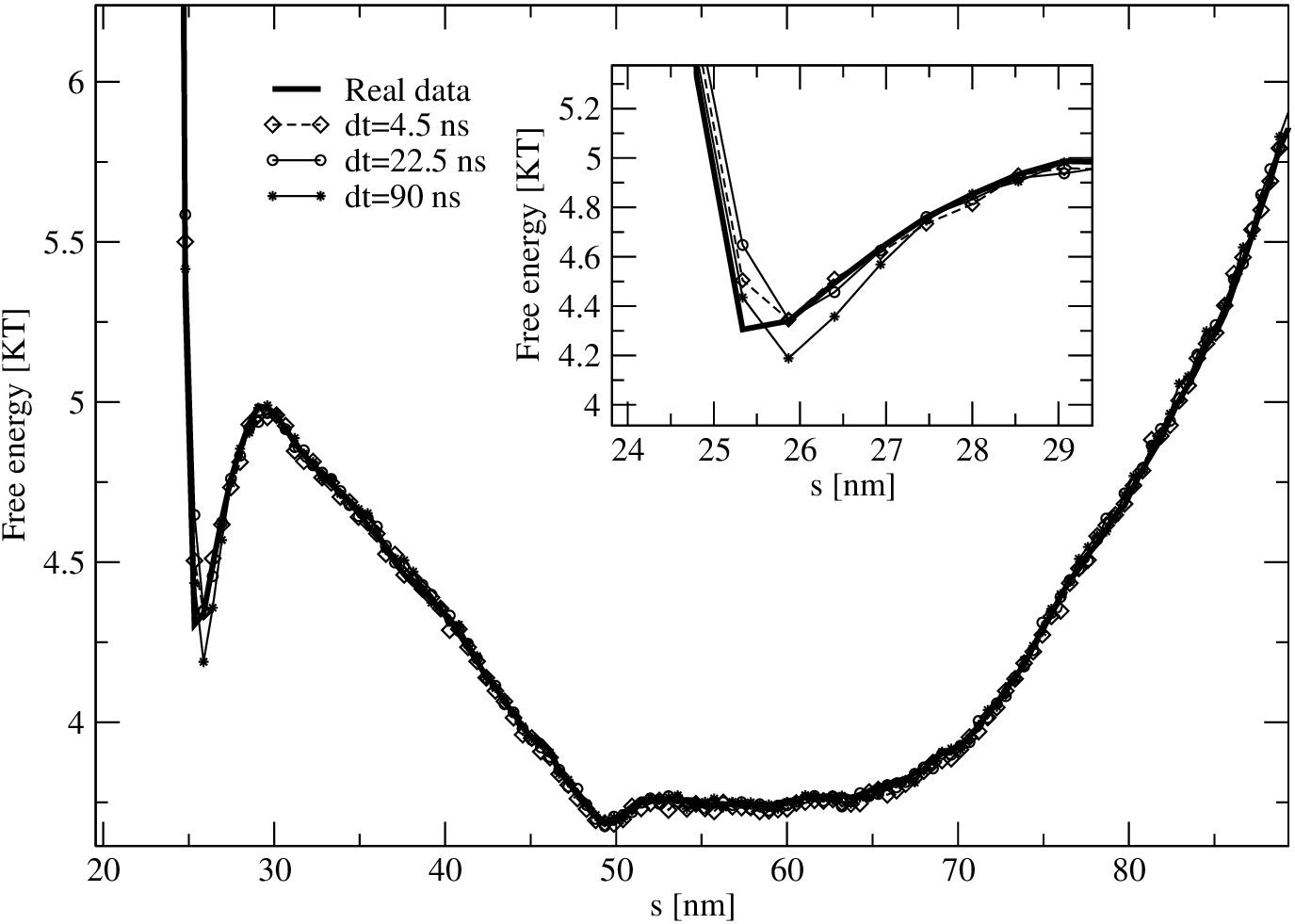}
\caption{
Free energy. Thick line: profile obtained from the the 180 ms
trajectory. Other curves: free energy profiles reconstructed from the
optimally determined $v(s)$ and $D(s)$, for various choices of $dt$. The
free energies are from the histogram of $s$ as as $F\left( s\right) =-\frac{1}{\protect\beta }\log N\left( s\right) =-\frac{1}{\protect\beta }\log
\protect\int dt\protect\delta \left( s-s\left( t\right) \right) $.
The different curves are not distinguishable on the scale of the figure.
Inset: zoom of the region of the first minimum, showing that small deviations
from the true profile are observed for $dt>50 ns$.
}
\label{fig:freeene}
\end{figure}

\noindent In Fig. \ref{fig:freeene}, we compare the ``true'' canonical free
energy of the system, $F(s)$, obtained through the histogram of the values
of $s$ recorded in a long trajectory, with the one obtained computing $D$
and $v$ by equations (\ref{eqn:solution1}), (\ref{eqn:solution2}), and
integrating numerically the stochastic differential equation formally given
in eqn. (\ref{langevin}). Specifically, the discrete-time evolution of $s$
implemented numerically was:

\begin{equation}
s(t+dt) = s(t) - v^*(s) dt_l + \sqrt{2 D^*(s) dt_l} \cdot \eta(t)
\label{eqn:modelLang}
\end{equation}

\noindent where $\eta (t)$ is drawn from a Gaussian distribution with
unit variance. In order to avoid systematic error deriving from a
finite integration time, the time increment $dt_{l}$ is much smaller
than $dt$.  Qualitatively, the free energy profiles in Fig.
\ref{fig:freeene} show two minima: one (denoted by $L$, for looped, in
the following), for $s<27$, corresponding to a state in which the two
ends of the polymer are in contact; the other (denoted by $U$, for
unlooped), for $s>27$, corresponding to a state in which the two ends
of the polymer are far. As we already remarked, $D$ and $v$ depend on
the choice of the $dt$. The different curves are obtained solving the
Langevin equation using $D$ and $v$ determined using with $dt=$4.5, 18
and 45 ns. All the choices lead to approximately the same profile. The
profile starts degrading only for dt $>90ns$ (data not shown).

 This provides an \emph{a
posteriori} demonstration that $D\left( s\right) $ and $v\left( s\right) $,
if computed with an appropriate choice of $dt$, are consistent with the true
equilibrium properties of the system.

\subsection{Kinetic properties of the optimal model}

\begin{figure}
\includegraphics*[width=0.45\textwidth]{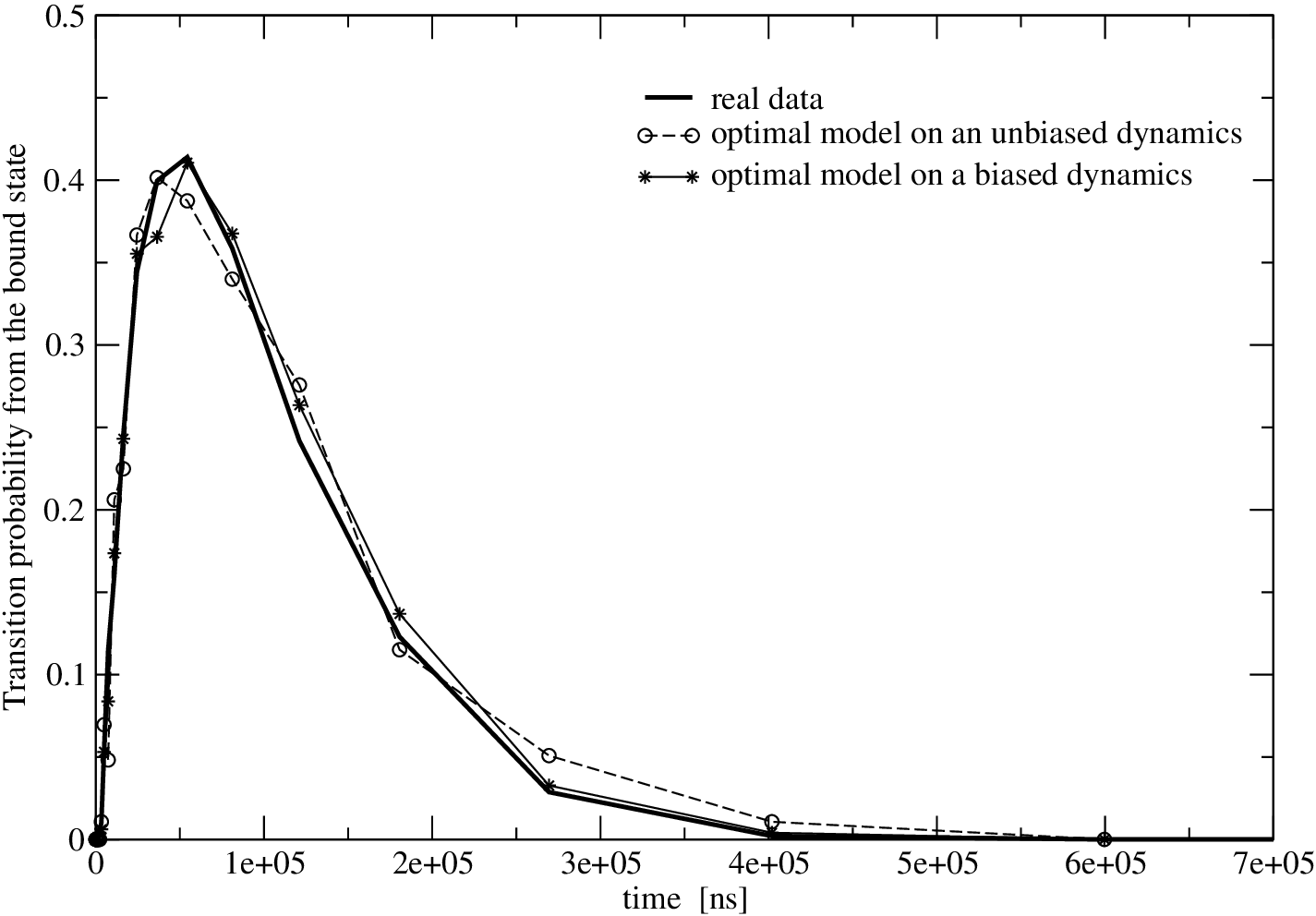}
\caption{
Normalized histogram of the residence times in the L state,
computed from the real dynamics and from the trajectories generated by the
optimal model.  The $L$ state corresponds to
whenever $s$ is smaller than 26, and enters the $U$ state whenever $s$ is
larger than 34.
}
\label{fig:tt}
\end{figure}

A complementary, stringent, test of the viability of the recovered
$v^{\ast } $ and $D^{\ast }$ profiles can be performed by
investigating kinetic-related properties of the system. The trajectory
obtained from the Langevin model (\ref{eqn:modelLang}) was processed
to calculate the average residence time in the bound state, namely the
time $\tau _{1}$ required by the system entering in the $L$ state to
escape from the well and entering in the $U$ state.  The normalised
distributions of $\tau _{1}$ obtained from the model Langevin equation
(\ref{eqn:modelLang}) and from the original trajectory are shown in
Fig. \ref{fig:tt}.

\noindent It can be seen that the two sets of distribution profiles are
very consistent, thereby indicating the viability of the model Langevin
description also for the kinetic system properties over time-scales much
larger that $dt$.

\subsection{Estimating the error}

\begin{figure}
\includegraphics*[width=0.45\textwidth]{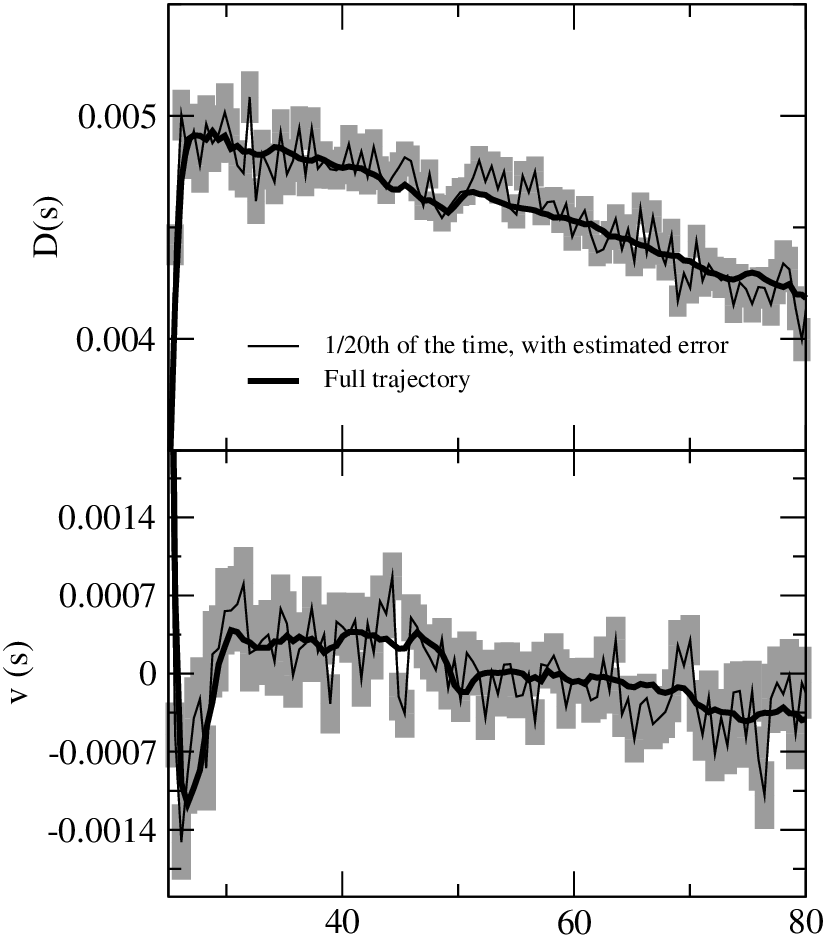}
\caption{
D(s) and v(s) evaluated with
  Eqs. (\protect\ref{eqn:solution1}) and (\protect\ref{eqn:solution2}) on
  an unbiased trajectory. Thick and thin lines are used for quantities
  obtained from trajectories of duration equal to 10ms and 0.5ms,
  respectively. The filled gray boxes represent the error bars
  calculated from Eqs. (\protect\ref{eqn:errors}) for the 0.5 ms-long case.
}
\label{fig:error}
\end{figure}

The validity of equation (\ref{eqn:errors}) for estimating the error has
been tested by comparing $D\left( s\right) $ and $v\left( s\right) $
computed from trajectories of different length. The results are shown in
Fig. \ref{fig:error}. The thick line represent $D\left( s\right) $ and $v\left( s\right) $ computed using all the 180 ms of the trajectory. As we
anticipated, even if the chosen collective variable is very simple, $D\left(
s\right)$ shows significant variations as a function of $s$. The thin black
line corresponds to the two quantities evaluated using a much shorter
trajectory of 9 ms. The solid gray blocks are the estimated errors as
given by Eqs. (\ref{eqn:errors}). The thin black lines
falls well within the estimated error of the 180 ms result, showing
that Eqs. (\ref{eqn:errors}) provide viable estimates for the statistical
uncertainties.

\subsection{{Optimal model of the non-externally-driven system from an out-of-equilibrium trajectory}}
\label{sec:dt}

A major advantage of the approach presented here is that it can be
applied also on trajectories generated under the action of an external
time-dependent forces. To illustrate this point we now consider a
trajectory of the system under the action of an external force of the
form
\begin{equation}
\theta \left( t\right) =-\frac{d}{ds}\left[ \frac{1}{2}k\left(
s-s_{rest}\left( t\right) \right) ^{2}\right] 
\end{equation}\noindent If this force is applied, the system is biased towards following $s_{rest}\left( t\right) $.

\begin{figure}
\includegraphics*[width=0.45\textwidth]{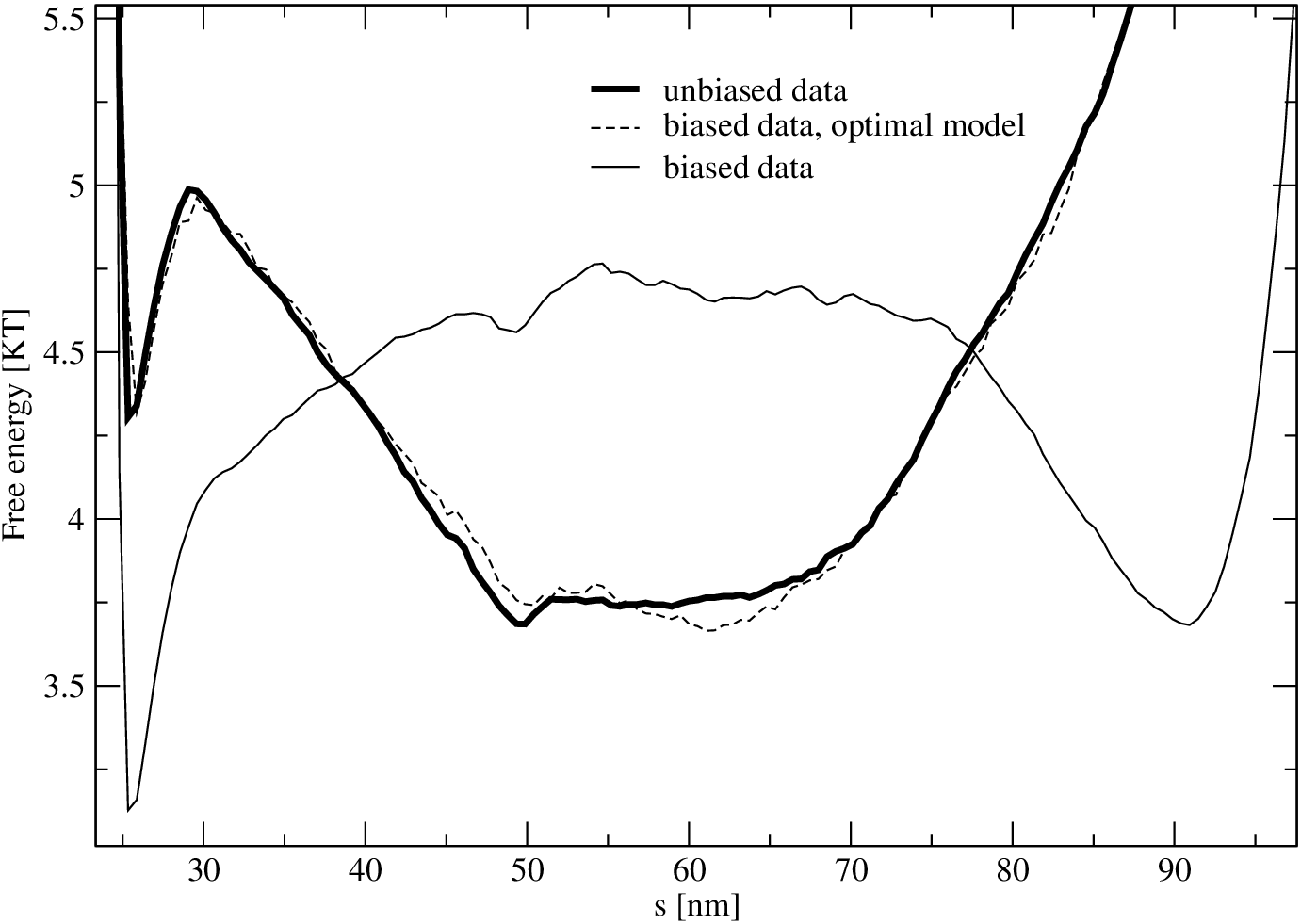}
\caption{
Free energy profiles. Thick line: {profile
obtained from the histogram of the trajectory of the
non-externally-driven system.} Dashed line: free energy profile
obtained from the optimal Langevin description applied to data
recorded in the presence of the harmonic time-dependent external
force. The continuous line provides, for comparison the
\textquotedblleft free-energy\textquotedblright\ profile obtained
directly from the histogram of $s$ recorded in the run subject to the
external force.
}
\label{fig:bias}
\end{figure}

Although the externally applied force can be chosen \emph{a priori} so
to optimize the statistical uncertainty on the $D$ and $v$ profiles we
shall consider the very simple case of an harmonic force derived from
a harmonic restraining potential whose center is scillating between
$s_{min}$ and $s_{max}$ with a period $T =1 ms $: \noindent
\begin{equation}
s_{rest}\left( t\right) =s_{min}+\frac{1}{2}\left( s_{max}-s_{min}\right)
\left( \cos \left( 2\pi \frac{t}{T}\right) -1\right)
\end{equation}
\noindent The values $s_{min}$ and $s_{max}$ are set equal to 21 and
104 nm, which hence cover a range of the original parameter space wide
enough to encompass both minima of the free energy. A sample
trajectory obtained under the action of this bias is shown in
Fig. \ref{fig:traj}, lower panel. As visible, the added external force
influences heavily the evolution of the system which, in fact,
exhibits a noisy harmonic modulation.
The external bias is so strong that a direct use of the the recorded
$s$ trajectory to compute the system free energy from the usual
histogramming procedure would lead to a completely wrong free energy
profile (shown with a thin continuous line in Fig. \ref{fig:bias}). By
contrast, the use of the optimal Langevin scheme derived above is very
effective in subtracting the effect of the bias and yield a free
energy profile that is entirely compatible with the true one.  Notice
that the bias subtraction does not exploit the knowledge of the
instantaneous values of the external force, but relies merely on the
knowledge of the time-averaged of the bias as a function of the
collective variable.

\subsection{{Validity of the Markovian approximation and optimal choice of dt}}

{As already mentioned, the choice of the time lag at
which the data are recorded, $dt$, is essential for the viability and
consistency of the proposed method, particularly regarding the
Markovian character of the chosen collective variable. If $dt$ is too
small, Eq. (\ref{langevin}) is not adequate for describing the
evolution of $s$, as the noise term (reflecting the influence of the
``integrated'' degrees of freedom) would have a sizable
autocorrelation time.} On the other hand, if the time lag is too
large, there would be prominent variations of the free energy and
diffusion coefficient evaluated for two ``consecutive'' positions,
$s(t)$ and $s(t+dt)$. This would invalidate the assumption, see
Eq. (\ref{eqn:modelLang}), that the force acting at time $t$ depends
only on the instantaneous position, $s(t)$.

{
If the recovered diffusion coefficient is constant in parameter space
and the underlying free-energy profile is harmonic, the Markovian character of the
collective variable can be established by verifying the exponential decay of its autocorrelation
function. More sophisticated procedures must be followed to compute the memory kernel
when the drift and diffusion terms do not have a
structure as simple as the one mentioned above\cite{kneller2001,grubmuller_diff}.
Here we show a series of simple quantitative tests that can be used to
assess the Markovian character of the collective variable on the time-scale defined by the
sampling interval $dt$. These tests can be easily used to find an optimal value of $dt$.
}

To this purpose we solve Eq. (\ref{langevin}) with respect to the
noise $dW_{j}\left( t\right) :$
\begin{equation}
dW_{j}\left( t\right) =\frac{1}{\sqrt{2}}D_{ij}^{-1/2}\left( ds_{i}\left(
t\right) -v_{i}\left( s\left( t\right) \right) dt\,\right)\ .   \label{noise}
\end{equation}
\noindent Using the estimates $v\left( s\right) $ and $D\left(
s\right) $ given by Eqs. (\ref{eqn:solution1}) and (\ref{eqn:solution2}),
one can evaluate $dW_{j}\left( t\right) $ along the trajectory
$\left\{ \dots ,s\left( t\right) ,s\left( t+dt\right) ,\dots \right\}
.$ It is readily seen that $dW_{j}\left(t\right) $ from
Eq. (\ref{noise}) satisfies:
\begin{eqnarray*}
\left\langle dW_{i}\left( t\right) \right\rangle  &=&0 \\
\left\langle dW_{i}\left( t\right) dW_{j}\left( t\right) \right\rangle 
&=&dt\delta _{ij}\ .
\end{eqnarray*}

\noindent Yet, the internal consistency of the procedure requires that
$dW_{j}\left( t\right)$ is uncorrelated at different times and that
its probability distribution is Gaussian. These two properties are not
enforced in the optimization procedure, and they can hence be used to
validate, {\em a posteriori} its applicability.

\begin{figure}
\includegraphics*[width=0.48\textwidth]{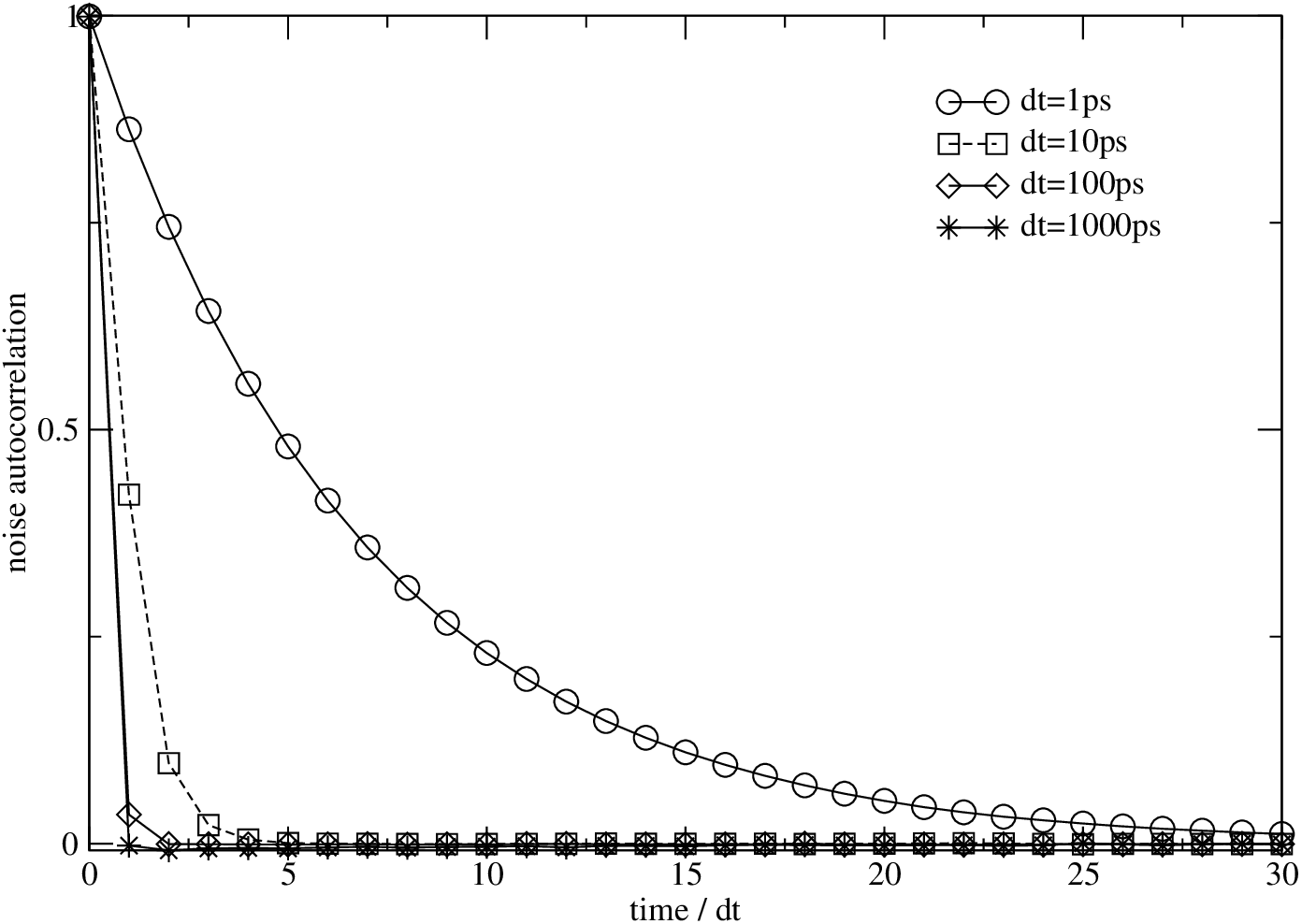}
\includegraphics*[width=0.48\textwidth]{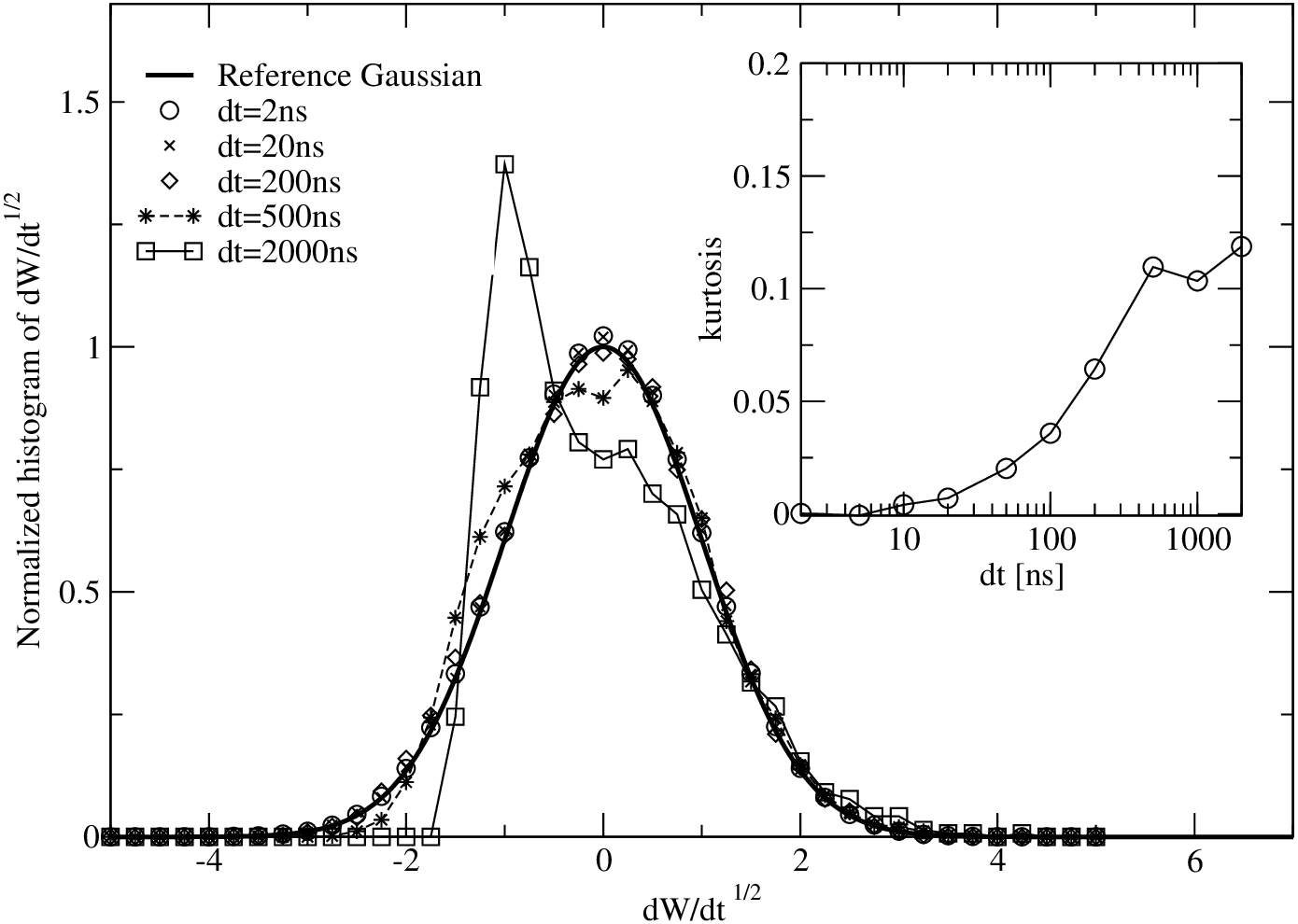}
\caption{
Upper panel: Time correlation function $C_{dt}\left( \tau
  \right) =\frac{1}{dt}\left\langle dW\left( t\right) dW\left( t+\tau
  \right) \right\rangle $ as a function of $\tau/dt$ for different
  choices of $dt$. Lower panel: Probability distibution of
  $dW/\sqrt{dt}$ for different choices of the time lag $dt$. $dW$ is
  computed from Eq.  \protect (\ref{noise}) and the average is
  restricted to values of $t$ in which $27.5<s(t)<30.5$ (approximately
  the region of the barrier).  Inset: normalized kurtosis, $\kappa
  \equiv (<dW^4> -3 <dW^2>^2)/dt^2 = <dW^4>/dt^2 -3$, as a function of
  the time lag $dt$. 
}
\label{fig:correlation}
\end{figure}

As a first step in the validation we calculate the autocorrelation
function of the noise for given values of the time lag $dt$,
$C_{dt}\left( \tau \right) =\frac{1}{dt}\left\langle dW\left( t\right)
dW\left( t+\tau \right) \right\rangle$. The averages were calculated
from a single underdamped Langevin evolution of the system and the
sampling time-lag, $dt$, ranged between 0.001 and 2000 ns. The resulting
autocorrelations as a function of $\tau/dt$ are shown in
Fig.~\ref{fig:correlation}, upper panel. The trends should be compared with the
step character of a memoryless noise: $C_{dt}\left( 0\right) =1$ and
$C_{dt}\left( \tau \right) =0$ for all $\tau \neq 0$. This limiting
behaviour is well-approximated for large $dt$, as the autocorrelation
drops almost immediately to zero ($C_{dt}\left( dt\right) \simeq 0.01$
for $dt=1$ ns). A slow decay is, instead observed for smaller $dt$'s,
indicating that the underlying stochastic process derives from a
correlated noise.

By inspecting suitable properties of the noise $dW_j\left( t\right)$
of Eq. (\ref{noise}) it is further possible to highlight the limitations of
excessively-large values of $dt$. A valuable indicator is provided by
the Gaussian character of the distribution of the instantaneous noise
amplitudes$\frac{dW\left( t\right) }{\sqrt{dt}}$. The histograms in
Fig. \ref{fig:correlation}, lower panel, indicate noticeable
deviations from Gaussianity for $dt$ larger than 500 ns. Also the
normalized kurtosis (see Fig. \ref{fig:correlation}, lower panel,
inset) becomes significantly different from zero for $dt>$ 100 ns.

It emerges that $dt$ must be chosen so to satisfy simultaneously the
criteria for the memoryless character of the noise and the Gaussianity
of its probability distribution.  For the specific system considered
here, it can be verified that using $dt$ between 1 and 100 ns provides
a viable and consisent Langevin description of the system. In fact, in
these conditions, the autocorrelation function of the noise has the
correct form and, at the same time, the probability distribution of
$dW$ is very close to a Gaussian.

\section{Conclusions}

We have presented an optimal scheme for describing {\em a posteriori}
the dynamics of a given system through the Langevin evolution of few
collective variables. The scheme lends to a straightforward numerical
implementation. It allows one to extract not only the drift profile
but also the diffusion coefficient which may both depend on the
collective variables. The proposed methodology allows to control the
statistical uncertainty affecting the calculated drift and diffusion
profiles. {Secondly, the drift and diffusion terms of
the non-externally-driven system can be recovered from trajectories
recorded in the presence of an externally-applied force.} This second
aspect appears particularly important as the external time-dependent
force can be designed to optimize, for a given duration of the system
evolution, the exploration of the phase space and control the
statistical uncertainty on the parameters of the Langevin
equation. The viability of the scheme was illustrated by applying it
to the looping kinetics of a model polymer system. As several other
approaches, the proposed one can be applied to systems describable by
an overdamped Langevin dynamics. We plan to explore the feasibility of
extending the present framework to the case of CV evolving under the
action of non-trivial memory kernels. This would be a particularly
important avenue for characterizing the salient dynamical features of
biomolecules which is presently attracting considerable attention due
to the large range of time scales that these molecules exhibit in
their internal dynamics
\cite{knel2000,hins2000,Frac_brown_2004,xie1,protG,karplus_nat2007b,NMRrotators}.

{\bf Acknowledgements} We thank M. Parrinello for valuable
discussions.  We acknowledge financial support from the Italian
Ministry for Education (FIRB 2003, grant RBNE03B8KK and PRIN, grant
2006025255) and from Regione Friuli Venezia Giulia (Biocheck, grant
200501977001).

\end{document}